\documentclass[%
 reprint,
 amsmath,amssymb,
 aps,
 prl
]{revtex4-1}

\usepackage{graphicx}
\usepackage{dcolumn}
\usepackage{bm}

\begin{document}


\title{Linking particle dynamics to local connectivity in colloidal gels}

\author{Jan Maarten van Doorn}
\author{Jochem Bronkhorst}
\author{Ruben Higler}
\author{Ties van de Laar}
\author{Joris Sprakel}%
 \email{joris.sprakel@wur.nl}
\affiliation{%
Physical Chemistry and Soft Matter, Wageningen University \& Research, Stippeneng 4, 6708 WE, Wageningen, the Netherlands
}%

\date{\today}

\begin{abstract}
Colloidal gels are a prototypical example of a heterogeneous network solid whose complex properties are governed by thermally-activated dynamics. In this Letter we experimentally establish the connection between the intermittent dynamics of individual particles and their local connectivity.  We interpret our experiments with a model that describes single-particle dynamics based on highly cooperative thermal debonding. The model, in quantitative agreement with experiments, provides a microscopic picture for the structural origin of dynamical heterogeneity in colloidal gels and sheds new light on the link between structure and the complex mechanics of these heterogeneous solids. 
\end{abstract}

\maketitle

Attractive interactions can drive a dilute colloidal suspension towards a solid state formed by a sample-spanning and mechanically-rigid particle network \cite{zaccarelli2007colloidal, trappe2004colloidal}. These colloidal gels are non-equilibrium solids, kinetically arrested en route to their equilibrium state of solid-liquid coexistence \cite{lu2008gelation}. Such particle gels are characterized by strong heterogeneity in their local connectivity, mesoscopic structure and their dynamics and mechanics \cite{duri2006length, dibble2008structural, gao2007direct, 0953-8984-14-33-303}. The microstructure and internal dynamics of colloidal gels can be directly observed with microscopy techniques at the single-particle level. As a consequence, it forms an interesting testing ground to explore the complex and length-scale dependent mechanics of heterogeneous solids. Colloidal gels derive their mechanical rigidity from physically bonded gel strands and nodes that form a percolating elastic network. The linear elasticity of gels is governed by the mechanics of the network architecture and its thermal fluctuations  \cite{PhysRevLett.80.778, rueb1997viscoelastic}. By contrast, the gradual aging of gels to a denser state \cite{cipelletti2000universal, zaccarelli2007colloidal} and their non-linear response to applied stresses \cite{PhysRevLett.106.248303, gibaud2016multiple}, is governed by events occuring at the the much smaller length scale of individual particles. Since the bonds between the particles are typically weak, single particles can debond from strands in the gel by thermally-activated bond breaking \cite{lindstrom2012structures}. On longer time scales, this result in the gradual restructuration of the gel network, causing it to coarsen, age and relax internal stresses that are built up during gelation \cite{negi2009dynamics}. Moreover, thermal-activation at the single particle level plays a crucial role in processes of fatigue that preempt stress-induced failure of the gel network \cite{PhysRevLett.106.248303}. To date, quantitative descriptions of these thermally-activated phenomena have relied on mean-field approximations\cite{lindstrom2012structures}. Yet, the inhomogeneity in local coordination that is intrinsic to gels, must play a large role in the intermittent debonding dynamics that are at the origin of this complex non-linear behavior. As a result, linking the structure of colloidal gels to their non-linear mechanics has remained challenging, in particular as the relationship between local connectivity and thermally-activated dynamics of single particles is not clearly established.\\ \indent In this letter we explore the connection between the local connectivity and intermittent bonding-debonding dynamics of individual particles in colloidal gels. We use quantitative three-dimensional microscopy to experimentally probe this relationship in colloidal gels formed from colloids that interact by means of  short-ranged attractions. We show how the experimental data can be quantitatively described with a  microscopic model that describes particle debonding as a strongly cooperative thermally-activated event depending on the local bonding structure. This allows us to explain how the complex ensemble-averaged mean-squared displacement results from the convolution of different particle species within a single gel. Our results illustrate how the the heterogeneous dynamics characteristics of strongly disordered solids emerge from their complex and inhomogeneous local network structure.\\
\indent We study gels formed from poly(methyl methacrylate) (PMMA) particles, stabilised by a poly(hydroxystearic acid) comb polymer, synthesized as detailed elsewhere \cite{antl1986preparation}. The particles have a radius $a = 709$ nm and a polydispersity of $\sim 5$\%, as determined from static light scattering experiments. The particles are equilbrated and suspended at a nominal volume fraction $\phi = 0.20$ in a density-matching solvent mixture of cyclohexyl bromide and decalin. The solvent is saturated with tetrabutylammonium bromide (TBAB) to partially screen charge interactions; we note that even at saturation, very weak electrostatic interactions remain \cite{C2SM26245B}.  Attractive forces between the particles are induced by the addition of polystyrene ($M_w =$ 105 kg/mol, $M_w/M_n =$ 1.06) as a depletant. In our solvent, this polymer has a radius of gyration $R_g \approx$ 10 nm, resulting in a short ranged depletion attraction with $\xi = R_g/a = 0.014$. Three-dimensional image stacks with a field-of-view of 41x41x21 $\mu$m$^3$ are acquired with confocal fluorescence microscopy at 1Hz; for each sample we capture 5000 stacks to ensure sufficient statistics. The three-dimensional centroid positions and the trajectories of all particles in the field-of-view are then determined with a resolution $dr_{res} = 40$ nm \cite{gao2009accurate}. \\
\begin{figure}
\centering
\includegraphics[width=\linewidth]{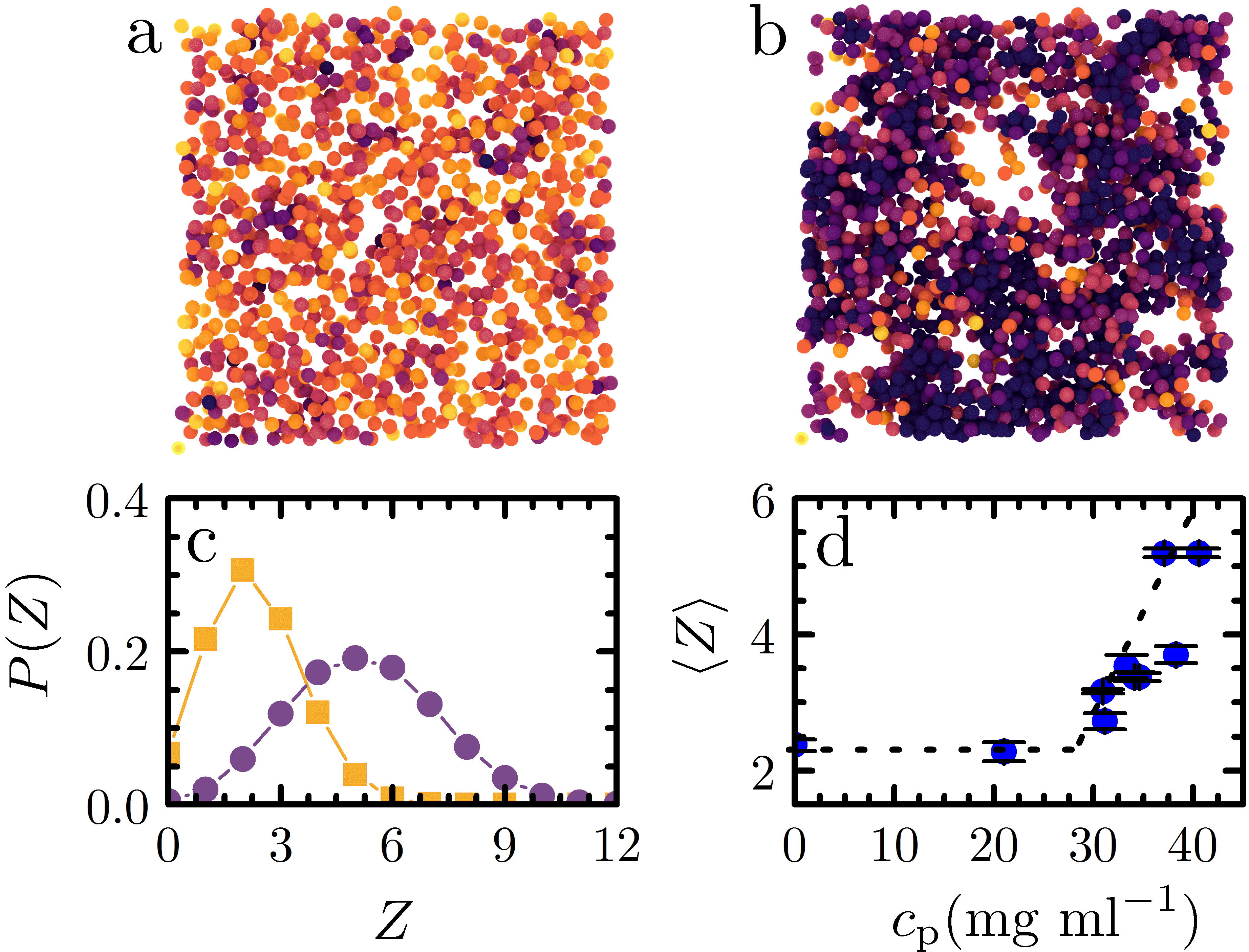} 
\caption{(color online) a-b) Computer-generated renderings of a liquid just before the gel point ($c_p = 21.0$ mg/ml) and a gel ($c_p = 37.1$ mg/ml)  based on three-dimensional confocal microscopy data, with particles color coded according to their coordination number (from dark blue $Z \geq 6$ to yellow $Z = 1$). c) Coordination number distributions for a liquid $c_p = 21.0$ mg/ml (squares) and a gel $c_p = 37.1$ mg/ml (circles). d) Ensemble-averaged coordination number $\langle Z \rangle$ as a function of depletant concentration $c_p$, dotted lines to guide the eye. }
\label{fig1}
\end{figure}
\indent Upon increasing the polymer concentration $c_p$ in a suspension of these particles, the structure of the sample transitions from a fluid of isolated particles, into a fluid of small and dynamic clusters \cite{lu2006fluids}. At a threshold depletant concentration a sample-spanning gel structure forms (Fig.~\ref{fig1}b). The phase behavior of this experimental system was studied in detail previously \cite{lu2008gelation, pusey1994phase}. To evaluate the sample microstructure, we first calculate the ensemble-averaged and static coordination number $\langle Z \rangle$ from snapshots of the three-dimensional gel structure. As the attraction strength increases we see a transition from a low, but finite, value of $\langle Z \rangle$ in the liquid state, and a rapid growth in the coordination number as the sample transforms into an aggregated colloidal gel (Fig.~\ref{fig1}d) \cite{dibble2008structural}. However, the average coordination number does not provide insight into the strong intrinsic heterogeneity in the microscture of colloidal gels, which becomes visible in a computer-generated representation of our experimental system in which the particles are color-coded according to their instantaneous value of $Z$ (Fig.~\ref{fig1}a-b). Indeed, calculation of the coordination number probability $P(Z)$ reveals a relatively wide distribution, both prior-to and beyond the gelation threshold (Fig.~\ref{fig1}c).\\
\indent The microscopic dynamics of colloidal systems are conventionally probed by means of the time- and ensemble-averaged mean-squared displacement (MSD) $\langle \Delta r^2 \rangle$ (Fig.~\ref{fig2}a). In these data, a continuous transition between fluid and solid behavior is observed. At low attraction strengths a diffusive $\langle \Delta r^2 \rangle \propto t$ is found (Fig.~\ref{fig2}a). Beyond a threshold $c_p \sim 30$ mg/ml,  $\langle \Delta r^2 \rangle$ decreases and begins to display a time-independent localisation plateau at short lag times. The height of this plateau $\delta^2$  decreases with increasing $c_p$, while it extends to increasingly large lag times. At even longer times $\langle \Delta r^2 \rangle$ exhibits an upturn to diffusive behavior. The value of $\langle \Delta r^2 \rangle$ at a lag time $t = 498$ s, as a proxy for the low-frequency particle mobility, exhibits a continuous transition between fluid-like behavior at low $c_p$ to a gel-like state for $c_p > 33$ g/L (Fig.~\ref{fig2}b). \\
\begin{figure}
\centering
\includegraphics[width=1.0\linewidth]{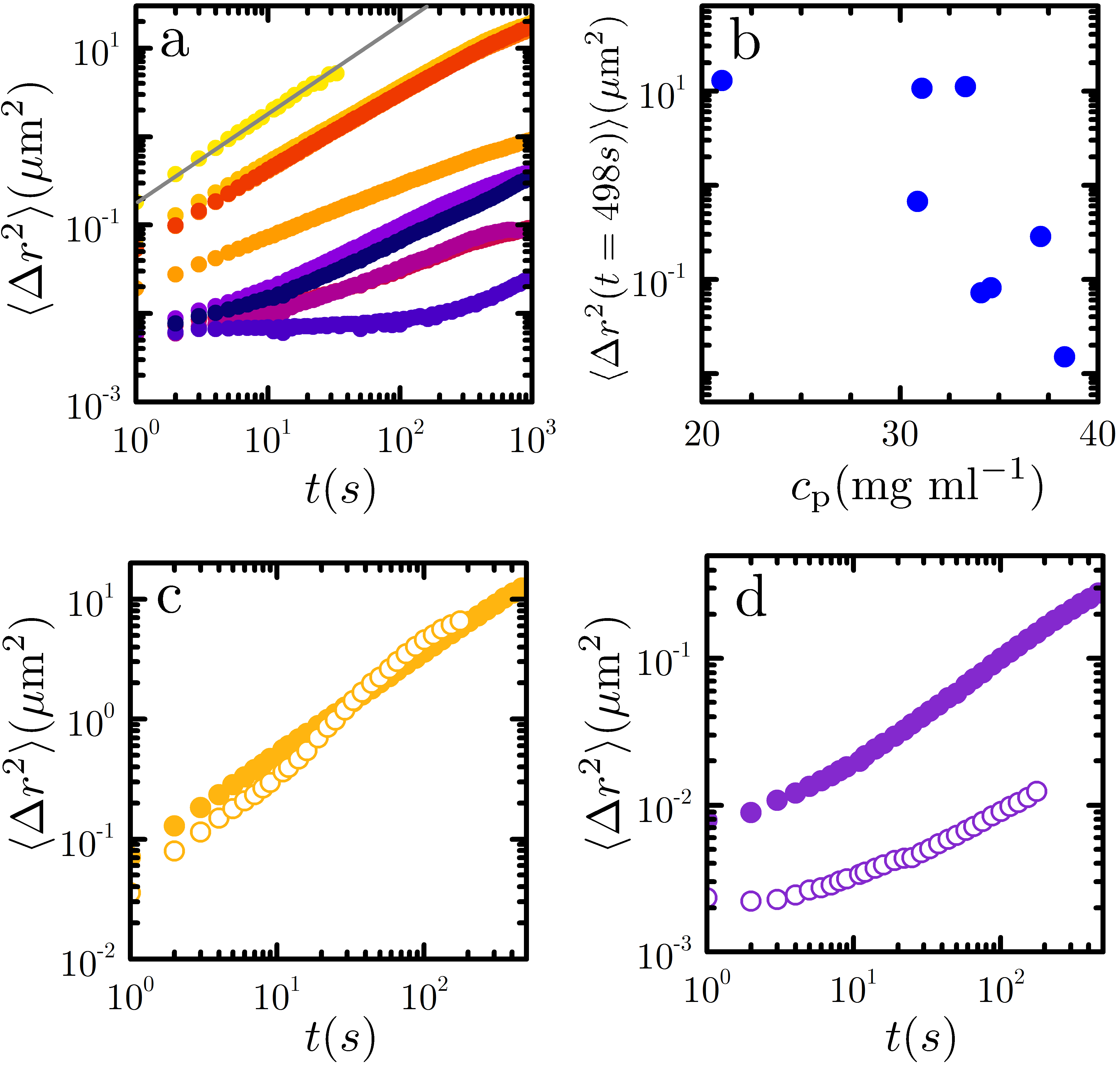} 
\caption{(color online) a) Ensemble- and time-averaged self-part of the mean squared displacements as a function of polymer volume fraction with (from top to bottom): $c_p = $0, 21.0, 30.9, 31.1, 33.3, 34.1, 34.6, 37.1, 38.3, 40.6 mg/ml. b) Value of $\langle \Delta r^2 \rangle$ at $t = 498$ s as a function of $c_p$. c-d) self (closed symbols) and distinct part (open symbols) of the mean-squared displacement for a sample in the liquid (c, $c_p = 21.0$ mg/ml) and in the gel (d, $c_p = 37.1$ mg/ml).}
\label{fig2}
\end{figure}
\indent The conventional, self-part of the MSD is a measure for the local dynamics of single particles. To illustrate the fact that the internal gel dynamics are strongly length-scale dependent, we compare these data to the distinct-part of the mean-squared displacements $\langle \Delta r^2 \rangle_D$ (Fig.~\ref{fig2}c-d). The distinct, or 2-point, mean-squared displacement, computed as described elsewhere \cite{crocker2000two}, probes the correlated motion of particles transmitted through the medium. As such they are a measure for the global, rather than local, properties of the gel. For samples in the fluid, just prior to the liquid-solid transition, the self- and distinct-parts of the MSD overlap within experimental error (Fig.~\ref{fig2}c). This indicates that there are no appreciable differences between local and global dynamics. By contrast, just above the gel threshold the distinct $\langle \Delta r^2 \rangle_D$ is almost an order-of-magnitude lower than the self-part of the MSD (Fig.~\ref{fig2}d). The gel is more rigid at the macroscopic scale, than that what is experienced by individual particles locally. Apparently, the dynamics of single particles in the gel are strongly affected by local structures; insight into these effects cannot be obtained by ensemble averaging.\\
\indent We hypothesize that single-particle dynamics, as measured by the self-part of the MSD, can be described by a specific sequence of events. Particles are first bonded to their neighbors in the gel network by bonds of strength $U/k_BT$. Under the action of thermal fluctuations, particles spontaneously debond from the gel with a characteristic rate $k_{d,Z}$; after debonding a particle will diffuse through the viscous medium with a rate $D$. This motion persists, until the particle collides with the gel network and re-attaches by forming new bonds. Thus, particles can exist in two states, bound and free, each characterised by different dynamics.\\
\indent We can experimentally evidence the existence of these two populations by determining the probability distribution $P(\Delta r^2 (t))$ of mean-squared displacement values for individual particles at a particular lag time $t = 498$ s. A sample in the fluid states exhibits a distribution with a single population of freely diffusing particles (Fig.~\ref{fig3}a), also illustrated by the linear dependence of the ensemble-averaged MSD with time (Fig.~\ref{fig2}a). By contrast, a sample in the gel state reveals two populations; a major fraction of the particles is bonded and exhibits a very low mobility, whereas a secondary peak signals the particles which have temporarily debonded and diffuse through the solution (Fig.~\ref{fig3}b). Note that this diffusive population has a lower effective diffusion coefficient that particles in the repulsive liquid, probably due to the fact that not only singlets, but also small clusters debond and diffuse. \\ 
\begin{figure}
\centering
\includegraphics[width=1.0\linewidth]{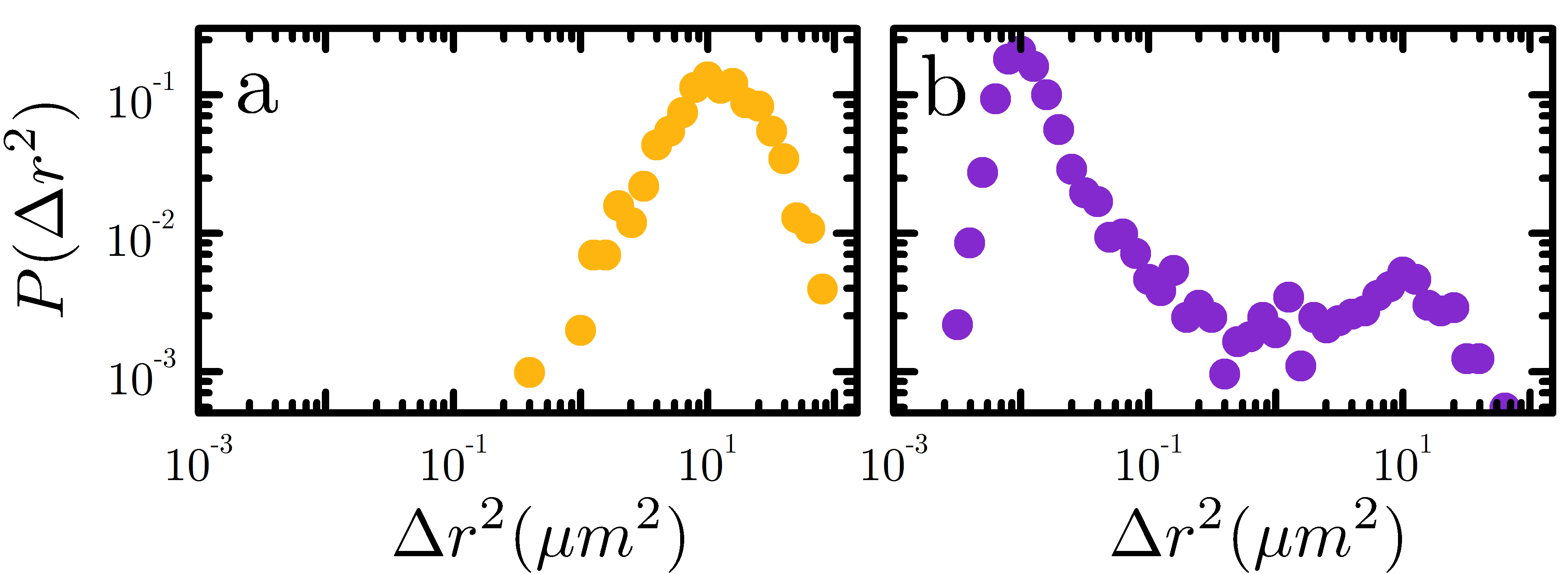} 
\caption{(color online) Probability distributions $P$ of single-particle mean-squared displacements at $t = 498$ s for $c_p = 21.0$ (a) and 37.1 mg/ml (b).}
\label{fig3}
\end{figure}
\indent The self-part of the mean-squared displacement of a single particle $\Delta r^2(t)$ can be split into two contributions: i) free diffusion during a characteristic time $\tau_{f}$, during which $\Delta r^2(t)=6Dt$ and ii) thermal vibrations of amplitude $\delta$ around an equilibrium bonded position, during a time $\tau_{b}$, for which $\Delta r^2(t) =\delta^2$. If we define $\alpha_{b} = {\tau_{b}} / {(\tau_{b} + \tau_{f})}$ as the fraction of time a particle resides in a bonded configuration, the time-averaged mean-squared displacement of a single particle can be approximated as:
\begin{equation}
\label{MSD}
\Delta r^2(t) = (1-\alpha_b) 6Dt + \alpha_b \delta^2
\end{equation}
\noindent For the sake of simplicity, we presume that the diffusion of debonded species occurs at a rate $D = ^{k_BT} / _{6\pi\eta a}$, where $\eta = 2$ mPa s is the viscosity of the suspending medium. \\
\indent The localization length $\delta$ of a bonded particle is set by the curvature of the local minimum in the potential energy. In our experiments we use depletion interactions; this gives rise to an attraction of depth $U$ and a range of the order of the depletant $R_g$. Approximating the bonds as a harmonic well $U = k \Delta r^2$, we estimate the spring constant of the bond between two colloids from dimensional analysis as $k \sim U / {R_g^2}$. Thermal excursions from their equilibrium position will occur with a typical squared amplitude $\delta^2 \sim {k_BT} / {k} = {k_BT R_g^2}/{U}$. Note that, in our experiments, only vibrations that exceed the spatial resolution $dr_{res}$ of the particle locating algorithm can be detected. Smaller vibrations will result in a observed mean-squared displacement plateau of $\delta^2 \approx dr_{res}^2$.\\
\indent The characteristic time a particle resides in a bound state is governed by thermally-activated dynamics. In an Eyring approach, the rate of dissociation of a single bond is described as $k_{d,1} = \omega_0 \exp \left[-U/k_BT\right]$, where $\omega_0$ is the attempt frequency \cite{eyring1935activated}. For a particle to detach from the gel network, all $Z$ bonds that connect it to its neighbors must be ruptured. Breaking one bond, while the particle stays in place due to the remaining $Z-1$ bonds, leads to rapid restoration of the broken bond with a rate $k_a$. Assuming that $k_a \gg k_{d,1}$, particle detachment from the network will only occur if all $Z$ bonds break simultaneously \cite{lindstrom2012structures}. Thus particle detachment is a strongly cooperative process with a rate $k_{d,Z} = (k_{d,1})^Z$. The typical time a particle remains bonded becomes $\tau_b = \frac{1}{\omega_0}\left[\exp\left({ZU}/{k_BT} \right)-1\right]$,where the term $-1$ ensures that the bonding time vanishes as $Z \rightarrow 0$. Substituting these results in Eq.\ref{MSD} gives a microscopic expression for the single-particle mean-squared displacement as:
\begin{equation}
\label{MSD2}
\Delta r^2(Z,t) = 6Dt + \frac{e^{ZU/k_BT}-1}{e^{ZU/k_BT}-1+\omega_0\tau_f} \left(\delta^2-6Dt \right)
\end{equation}
\noindent This expression predicts a distinct dependence of the single-particle dynamics on its local coordination number $Z$. From our experimental data, we determine the value of $\Delta r^2 (Z,t)$ at a fixed lag time $t = 498$s, and plot these as a function of the average coordination number for the particle during the length of our experimental observations (symbols in Fig.~\ref{fig5}a). We fit these experimental data to the theoretical model (Eq.~\ref{MSD2}), in which there are two fitting parameters: the effective energy of interparticle bonds $U$ and the dimensionless number $\omega_0 \tau_f$, which is the ratio of the frequencies of debonding attempts and reassociations. The predictions from the microscopic theory are in excellent agreement with the experimental data (symbols in Fig.~\ref{fig5}a). Both data sets, for different polymer concentrations, can be fitted with $\omega_o \tau_f \approx 0.1$, which indicates that particle reassociation is indeed significantly faster than debonding, thus confirming the validity of the assumption that $k_a \gg k_{d,1}$. 

The effective bonding energies we need to fit the data in proximity to the gel point are of the order of  $\sim$ 5 $k_BT$; these values are almost an order-of-magnitude lower than the depth of the depletion attraction calculated with the Asakura-Oosawa model \cite{asakura1958interaction}, which assumes only hard sphere repulsions. We attribute this to the still significant electrostatic repulsion known to act between PMMA particles in apolar solvents even in presence of the TBAB electrolyte \cite{C2SM26245B}.

\begin{figure}
\centering
\includegraphics[width=1.0\linewidth]{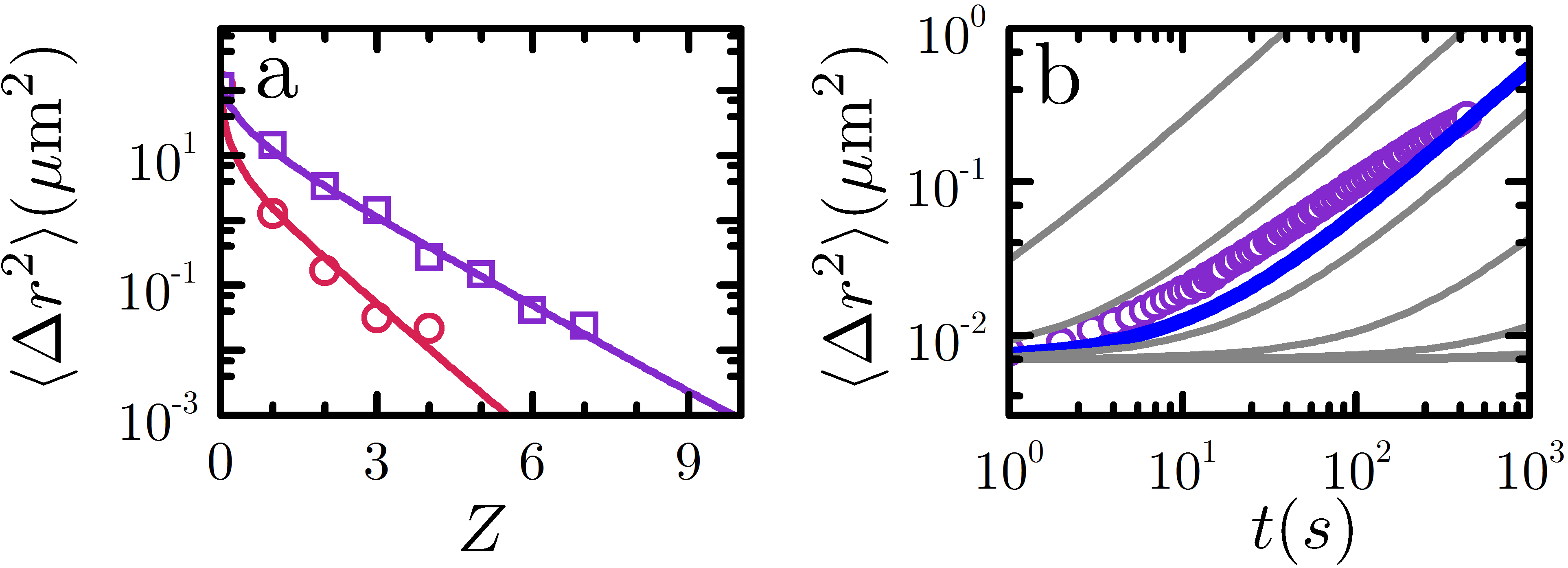} 
\caption{(color online) a) Single-particle mean-squared displacement at $t = 498$ s as a function of particle connectivity $Z$ from experimental data (symbols) and as predicted by the model described in the text (solid lines) for $c_p = $ 34.1 (circles) and 37.1 mg/ml (squares). b) Comparison between experimental ensemble-averaged $\langle \Delta r^2 \rangle$ (symbols) and that predicted by Equation~\ref{eq3} without adjustable parameters (solid blue line). Solid gray lines are the contributions to the mean-squared displacements as a function of particle coordination number with (top-to-bottom) $Z = $ 0, 2, 4, 6, 8, 10, as predicted by Eq.~\ref{MSD2}.}
\label{fig5}
\end{figure}

\begin{figure}
\centering
\includegraphics[width=1.0\linewidth]{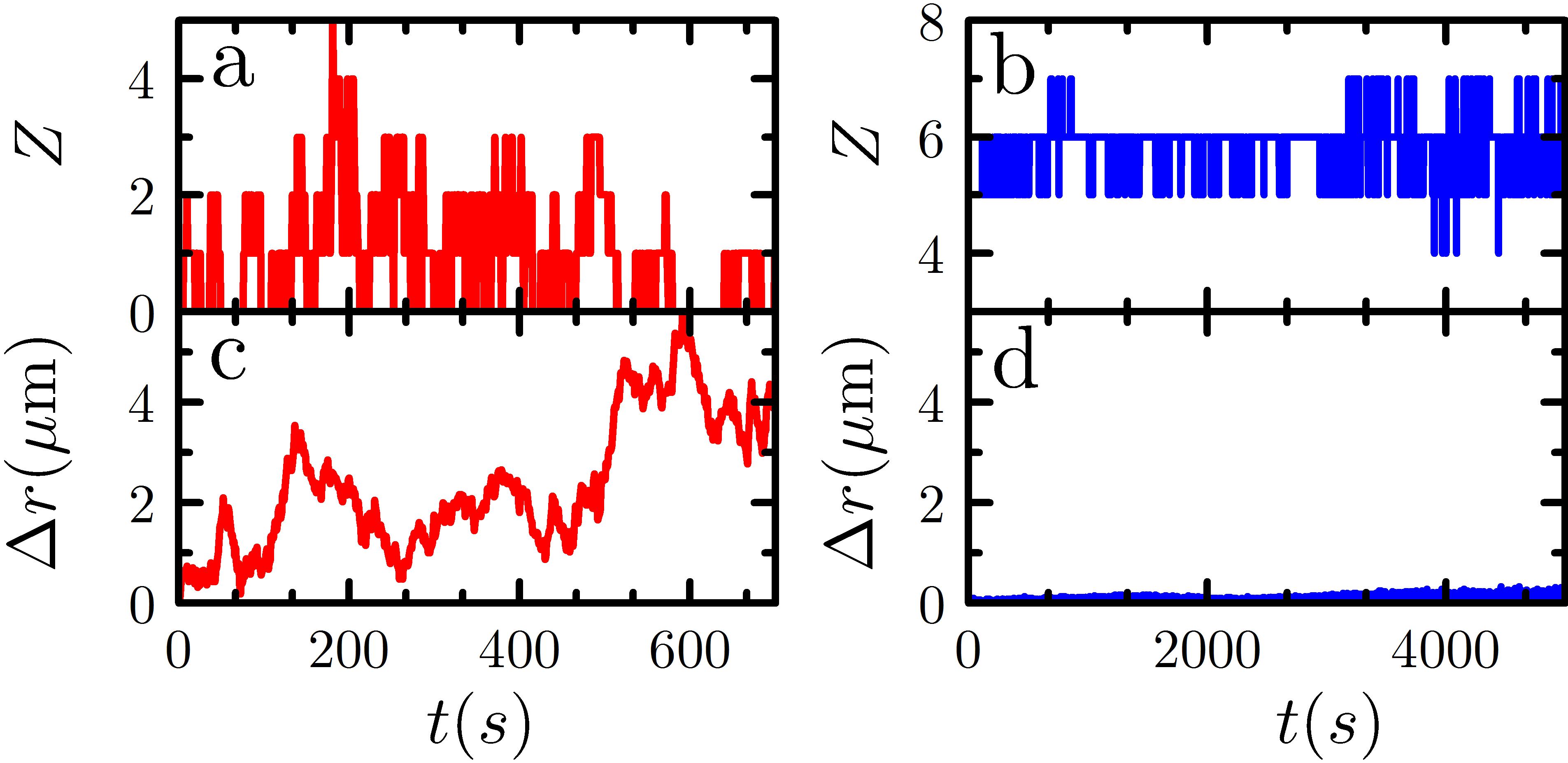} 
\caption{(color online) Thermally-activated fluctuations in the coordination number $Z$ of a single particle (a,c) and the corresponding particle displacement $\Delta r$ (b,d) for a weakly connected (a-b) and highly connected (c-d) particle in the same gel at $c_p = 37.1$ mg/ml. Note that the trajectory length is much shorter for the weakly connected particles as it diffuses out of the field-of-view after $\sim 700$ s.}
\label{fig4}
\end{figure}

These data illustrate the intimate link between single-particle dynamics and local connectivity. To further substantiate these findings we probe the evolution of the coordination number for a single particle as a function of time. For a weakly connected particle, strongly intermittent fluctuations occur between bonded $Z>0$ and unbonded $Z=0$ states (Fig.~\ref{fig4}a); the continuous debonding and diffusion allows the particle to travel significant distances over the course of several minutes before it exits the field-of-view (Fig.~\ref{fig4}c). By contrast, a strongly coordinated particle shows fluctuations in coordination number of $\pm 1$ (Fig.~\ref{fig4}b), but remains connected over the entire length of the experiment of 5000s, and as a consequence only exhibits strongly localised positional fluctuations (Fig.~\ref{fig4}d).\\
\indent Finally, with a quantitative microscopic description for the effect of connectivity on single-particle dynamics (Eq.\ref{MSD2}), we attempt to reconstruct the ensemble-averaged mean-squared displacement. To do so, we must weight the ensemble-average using the distribution of coordination numbers $P(Z)$ as a weighting function: 
\begin{equation}
\label{eq3}
    \langle \Delta r^2 (t) \rangle  = \sum_Z P(Z) \Delta r^2(Z,t)
\end{equation}
\noindent With the values of $U$ and $\omega_0 \tau_f$ determined from our experimental data (Fig.~\ref{fig5}a) and  $P(Z)$ obtained directly from the static structure of the gel (Fig.~\ref{fig1}c), we can now predict the ensemble-averaged MSD. Indeed, without adjustable parameters, we find that the reconstructed $\langle \Delta r^2 (t) \rangle$ based on our model for single particle dynamics is in reasonable quantitative agreement with the ensemble-averaged MSD determined directly from experiments (Fig.\ref{fig5}b). This highlights the self-consistency of our description. Moreover, it enables us to deconvolve the ensemble-average into the different populations of particles with different local coordination numbers $Z$ (solid gray lines, Fig.\ref{fig5}b). This provides a direct and quantitative explanation for the distinct dynamical heterogeneities characteristic of colloidal gels. \\
\indent We have presented experimental data and theoretical analysis that explains how the heterogeneous dynamics of colloidal gels derives from the large inhomogeneities in local connectivity. The quantitative description of single-particle dynamics based on the local structure could form a stepping stone to develop microscopic descriptions of processes, such as aging, syneresis or stress-induced fatigue, in which the local microstructure evolves over time under the action of thermally-actived particle rearrangements. In our current description, we have only considered particle rearrangements to occur through debonding and reassociation onto the gel network. Even though this provides a reasonable approximation, given the agreement between our experiments and the model,  other thermally-actived modes of particle motion, such as the sliding of a particle along a gel strand without debonding entirely may exist. Increasing the attraction range, will make these types of rearrangements more likely to occur; extending our model to account for these "sliders", could lead to a more generalized descripion of local dynamics that is applicable to a wide range of disordered network materials, even those in which the local connectivity must be preserved \cite{montarnal2011silica}. 
\section*{Acknowledgements}This work is part of an Industrial Partnership Programme of the Foundation for Fundamental Research on Matter (FOM), which is part of the Netherlands Organisation for Scientific Research. The work of TvdL is carried out as part of a project of the Institute for Sustainable Process Technology: Produced Water
Treatment (WP-20-03).

\section*{References}

\bibliography{refs}

\end{document}